\documentclass[9pt,conference]{IEEEtran}

\usepackage{csquotes}
\usepackage{textgreek}
\usepackage{tabularx}
\usepackage{multirow}
\usepackage{multicol}
\usepackage{soul}
\usepackage{graphicx} 

\usepackage{dblfloatfix}    
\usepackage{textcomp}

\usepackage{booktabs} 
\usepackage{graphicx}
\graphicspath{{figs/}}
\usepackage{braket}
\usepackage{multirow}

\usepackage{cite}
\usepackage{amsmath,amssymb,amsfonts}
\usepackage{algorithmic}
\usepackage{graphicx}
\usepackage{textcomp}
\usepackage{xcolor}
\usepackage{color}
\def\BibTeX{{\rm B\kern-.05em{\sc i\kern-.025em b}\kern-.08em
    T\kern-.1667em\lower.7ex\hbox{E}\kern-.125emX}}

\begin{document}


\title{Analysis of Quantum Approximate Optimization Algorithm under 
 Realistic 
Noise in Superconducting Qubits\vspace{-0em}}

\author{
      Mahabubul Alam, Abdullah Ash-Saki, Swaroop Ghosh \\
  	\begin{tabular}{c}	
     {\normalsize Department of Electrical Engineering} \\
     {\normalsize Pennsylvania State University, University Park, PA-16802} \\
     {\normalsize mxa890@psu.edu, axs1251@psu.edu, szg212@psu.edu} \\
  	\end{tabular}  
  	}

\maketitle

\begin{abstract}
The quantum approximate optimization algorithm (QAOA) is a promising quantum-classical hybrid technique to solve combinatorial optimization problems in near-term gate-based noisy quantum devices. In QAOA, the objective is a function of the quantum state, which itself is a function of the gate parameters of a multi-level parameterized quantum circuit (PQC). A classical optimizer varies the continuous gate parameters to generate distributions (quantum state) with significant support to the optimal solution. Even at the lowest circuit depth, QAOA offers non-trivial provable performance guarantee which is expected to increase with the circuit depth. However, existing analysis fails to consider non-idealities in the qubit quality i.e., short lifetime and imperfect gate operations in realistic quantum hardware. In this article, we investigate the impact of various noise sources on the performance of QAOA both in simulation and on a real quantum computer from IBM. Our analyses indicate that optimal number of stages (p-value) for any QAOA instance is limited by the noise characteristics (gate error, coherence time, etc.) of the target hardware as opposed to the current perception that higher-depth QAOA will provide monotonically better performance for a given problem compared to the low-depth implementations.
\end{abstract}


\section{Introduction}

Quantum computing technology is getting traction. On one front, researchers are developing new hardware technologies to realize a qubit, the building block of a quantum computer. On the other side, new quantum algorithms are proposed to harness the unique properties of quantum computers to solve a certain class of classically intractable problems. We now have prototypical near-term quantum computers often termed as noisy-intermediate-scale-quantum (NISQ) computers. The NISQ devices have a limited number of qubits, and cannot handle quantum algorithms like Shor's algorithm which typifies the quantum computing paradigm at a practical scale. However, on a journey to prove \textit{quantum supremacy}, and to make best use of the available near-term devices, a number of quantum-classical hybrid algorithms \cite{farhi2014quantum, kandala2017hardware, romero2017quantum, dallaire2018quantum} based on variational principles have been proposed. In these hybrid algorithms, a quantum computer and a classical computer works in tandem to speed-up a problem over its completely classical version. A quantum processor prepares a quantum state using a parameterized quantum circuit (PQC) (a PQC is a quantum circuit consisting of parameterized gates). A repeated measurement of the quantum state generates an output distribution which is then fed to a classical optimizer. Based on the output distribution, the classical computer generates a new set of optimized parameters for the PQC which is then fed-back to the quantum computer. The whole process continues in a closed loop until a classical optimization goal is satisfied.
On the forefront of these class of algorithms is the quantum approximate optimization algorithm (QAOA) \cite{farhi2014quantum} which can address combinatorial optimization problems. 

In QAOA, the quantum state is prepared by a p-level variational circuit specified by 2p variational parameters. 
According to \cite{farhi2014quantum}, QAOA offers non-trivial provable performance guarantees even at the lowest circuit depth (p = 1), and the performance is expected to improve with the p-value \cite{zhou2018quantum}. 
The classical optimizer that is used to find the optimal variational parameters can also have an impact on the performance, and there is no general consensus on the best classical optimization algorithm for such hybrid techniques. A number of recent advances in finding good parameters have been made for QAOA  \cite{zhou2018quantum, wecker2016training, guerreschi2018qaoa, zhou2018quantum, crooks2018performance, guerreschi2017practical}. 
In \cite{brandao2018fixed}, the authors showed that typical QAOA instances have similar values for similar control parameters, and proposed reusing optimal parameters between similar problems. 
Zhou et al. also showed the optimal control parameters for similar QAOA instances have small variations between themselves \cite{zhou2018quantum}.

However, the performance of QAOA on practical quantum hardware with non-idealities (i.e., noises) is largely unexplored. The gate operations in near-term devices are inaccurate (gate error) that introduces error in the computation, and the qubits have a short life time (decoherence) i.e., qubits tend to lose the saved state spontaneously with time. Therefore, the parameter and the solution landscape for QAOA in noisy hardware can potentially be different from their ideal counterparts. 
Moreover, the performance gain with higher p-values is most likely to be limited as each added level introduces more gate error, and the circuit execution needs more time to complete which may exceed qubit life-time (coherence time). 
In this paper, we analyze the performance of the QAOA under realistic noise, and show performance bound with depth.

\begin{figure} [!ht] 
\vspace{-1em}
 \begin{center}
    \includegraphics[width=0.45\textwidth]{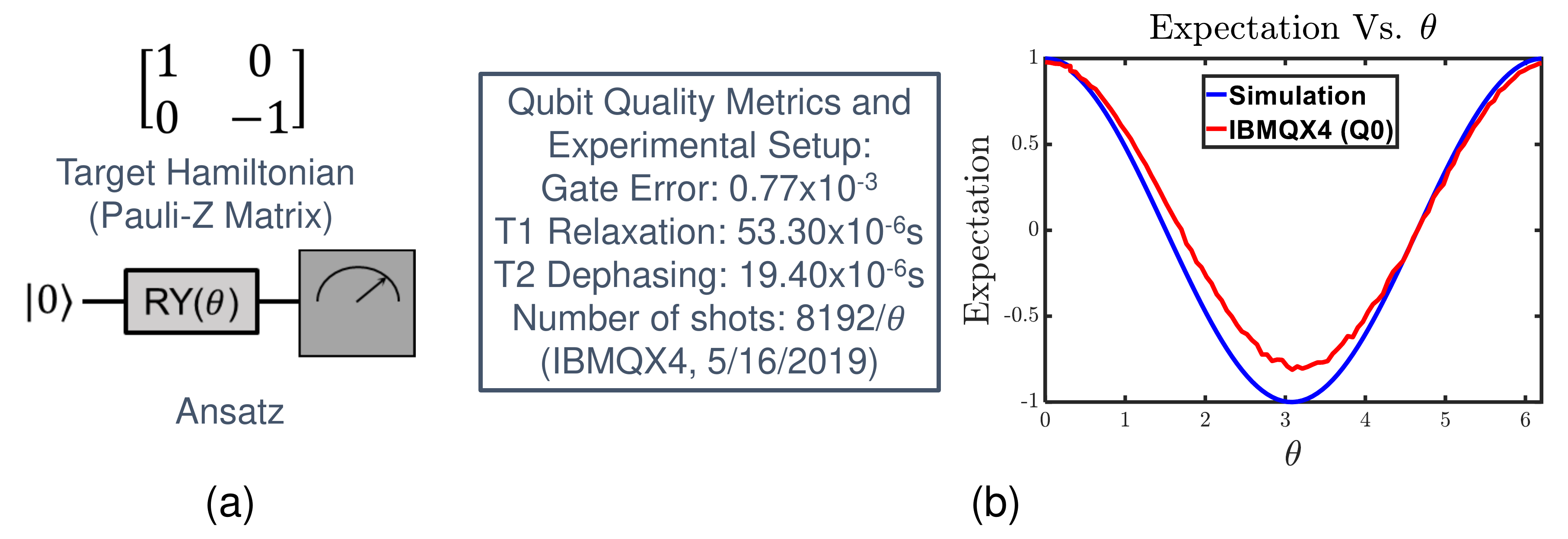}
 \end{center}
 \vspace{-1em}
 \caption{(a) Target Hamiltonian and a variational quantum circuit (Ansatz) constructed with a quantum gate with a single rotation parameter. The goal is to prepare an output state such that the expectation value of the Hamiltonian is $-1$; (b) experimental setup and the entire solution space of the circuit for noisy (IBMQX4) and noiseless quantum systems (simulation).} \label{fig:mot}
 \vspace{-0.5em}
\end{figure}

\textbf{Motivation:}
As a demonstrative example, we have executed a variational circuit (also known as a quantum ansatz) as in Fig. \ref{fig:mot}(a) on a simulator and a real device from IBM (IBMQX4). The target is to prepare an output state $\ket{\psi}$ by varying the $RY$ gate parameter $\theta$ such that the expectation value of Pauli-Z $(\sigma_z)$ operator in that state is the minimum i.e., $-1$ (such routine is often necessary in computing lowest eigenvalue). 
We have taken 100 samples of the expectation value in the entire solution space (0 to 2$\pi$) at equal distances with a simulator (without noise) and the target solution state $\ket{\psi} = \ket{1}$ is found at $\theta$ = $\pi$ which results in an expectation value of $-1$ as shown in Figure \ref{fig:mot}(b).
However, in our practical experiment with a real noisy hardware (IBMQX4) where the expectation value calculation can be affected by device imperfections like gate error, decoherence, etc., we have found that the exact solution ($\ket{\psi} = \ket{1}$ or Expectation value = $-1$) does not exist in the entire solution space.

As QAOA also works based on variational principles where we also search for feasible solutions in a quantum Hilbert space (i.e., search for a $\ket{\psi}$), the algorithm itself can be affected by the non-idealities of practical quantum hardware just like variational circuit in this example.  

\textbf{Contributions:}
In this article, we investigate the performance of QAOA in solving a combinatorial optimization problem (MaxCut) with the noise characteristics of a practical quantum computer (IBMQX4). We (a) demonstrate the steps of implementing QAOA, and summarize the associated challenges, (b) present our findings on the impact of different noise sources (i. relaxation, ii. dephasing, and iii. gate errors) on QAOA performance, especially with increasing p-values (higher-depth QAOA instances).  
To the best of our knowledge, this is the first work to analyze the QAOA performance with realistic noises of superconducting qubits - both analytically and experimentally. \textit{While claims from previous works indicates that the performance of QAOA will monotonically increase with a higher value of p, inciting optimism about the use case of near-term quantum devices, this paper shows that noise sources put a bound on those claims and projected improvement.} 

\section{Quantum Computing Basics and Analysis Setup}

\subsection{Quantum Computing Preliminaries}

\subsubsection{Quantum state and quantum gate}
In quantum computing, data is the state of a qubit, and computation is the quantum gate, which modulates the data or qubit state. Mathematically, a qubit state is generally represented using a state vector $\ket{\psi} = a \ket{0} + b \ket{1}$ such that $|a|^2 + |b|^2 = 1$. $\ket{0}$ and $\ket{1}$ are known as the computation basis states, and are expressed using vectors $[1, 0]^T$ and $[0, 1]^T$ respectively. 
Alternately, a qubit state is represented using density matrix $(\rho)$ such that $\rho = \sum_i p_i \ket{\psi_i} \bra{\psi_i} $ where $p_i$ = probability of pure state $\ket{\psi_i}$ in the density matrix. 

\begin{figure} [!hb]
\vspace{-1em}
 \begin{center}
    \includegraphics[width=0.45\textwidth]{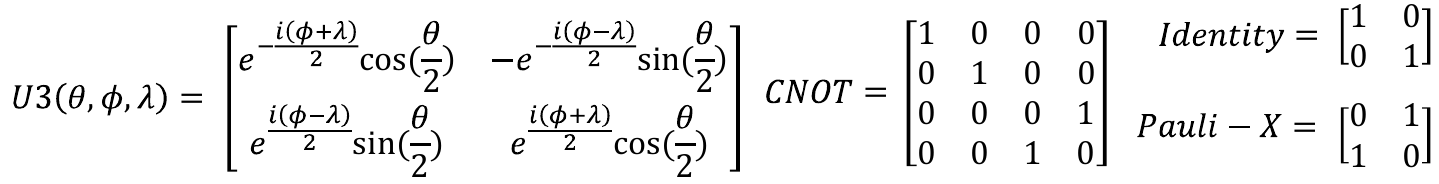}
 \end{center}
 \vspace{-1em}
 \caption{Gate matrices of the quantum gates used in this article.} \label{fig:matrix}
\end{figure}

A quantum gate is represented using unitary matrix known as the gate matrix. 
Quantum gates can work on a single qubit (e.g., Pauli-X ($\sigma_x$) gate) or on multiple qubits (e.g., 2-qubit CNOT gate). 
The gate matrices of the quantum gates used in this work are shown in Fig. \ref{fig:matrix}.

The computation on data (i.e., qubit state) is abstracted using matrix multiplication between density matrix and unitary gate matrix. In more generic terms, any operation $\mathcal{E}()$ that transform input $\rho_{in}$ to an output $\rho_{out}$ can be calculated using the operator-sum representation such that $\rho_{out} = \mathcal{E}(\rho_{in}) = \sum_k E_k \rho_{in} E_{k}^\dagger$, where $E_k$ matrix depends on the type of the operation. For example, if the operation is a single gate (say, $U(\theta, \phi, \lambda)$) on a qubit then $k = 1$ and $E_k = U(\theta, \phi, \lambda)$, and $\rho_{out} = U \rho_{in} U^\dagger$.

\subsubsection{Expectation Value}
Mathematically, the expectation value of an operator H in any state $\ket{\psi}$ is calculated as $\langle E \rangle = \bra{\psi} H \ket{\psi}$. In quantum computers, a qubit is measured in the so-called Z-basis or computational basis $\ket{0}$ and $\ket{1}$. These are the eigenvectors (eigenstates) of Pauli-Z ($\sigma_z$) operator with eigenvalues +1 and -1 respectively. Suppose the state of a qubit after a quantum computation routine is $\ket{\psi} = 0.8\ket{0} + 0.6\ket{1}$ (note the higher amplitude of $\ket{0}$). The expectation value of Pauli-Z operator in this state $\ket{\psi}$ is $\bra{\psi} \sigma_z \ket{\psi}$ = 0.28. Intuitively, the expectation is an single valued indication of the quantum state. The QAOA in this paper tries to maximize the expectation value of a specific cost Hamiltonian, thus, implicitly searches for a solution quantum state.

\subsection{Simulation and experimental setup}
To simulate the QAOA under noise, we use the noisy quantum computation simulator presented in \cite{ash2019qure} with the reported noise data from IBMQX4.  In the model gate error, relaxation, and dephasing are modeled with depolarizing, amplitude damping, and phase damping channel respectively. Input density matrix first undergoes ideal gate-operation followed by gate error, relaxation, and dephasing. To compute the output density matrix at each step, operator-sum method is used. The simulation flow finally gives a numerical density matrix incorporating the noise effects.

We also present experimental results by executing experiments on IBM's real quantum computer IBMQX4 \cite{IBMQ} (see Section \ref{sec:hw}). 

\section{Quantum Approximate Optimization Algorithm}

\subsection{Combinatorial Optimization}
Combinatorial optimization can be defined as the process of searching for maxima of an objective function whose domain is a discrete but large configuration space. Any combinatorial optimization problem can be defined on N-bit binary strings $z$ = \{$z_1$,$z_2$,....,$z_N$\}, where the goal is to determine a string that maximizes a given classical objective function $C$($z$): \{+1,-1\}$^N$ $\rightarrow$ $\mathbb{R}$ $_{\geq 0}$ ($z_i$ denotes a binary variable with two possible values: +1 or -1). The objective function has $m$ clauses and each of these clauses is a constraint of the bits which is satisfied for certain assignments of those bits and unsatisfied for other assignments and can be defined as $C$($z$) = $\sum_{\alpha = 1}^{m}$ $C_\alpha$($z$). Here, $C_{\alpha}$($z$) = 1 if $z$ satisfies clause $\alpha$ and 0 otherwise. Satisfiability asks if there is a string that satisfies every clause. MaxSat asks for a string that maximizes the number of satisfied clauses. An approximate optimization algorithm aims to find a string that achieves a desired approximation ratio $\frac{C(z)}{C_{max}}$ $\geq$ $r^*$ where $C_{max}$ = MaxSat($C$($z$)).

\begin{figure} [!ht] 
\vspace{-1em}
 \begin{center}
    \includegraphics[width=0.43\textwidth]{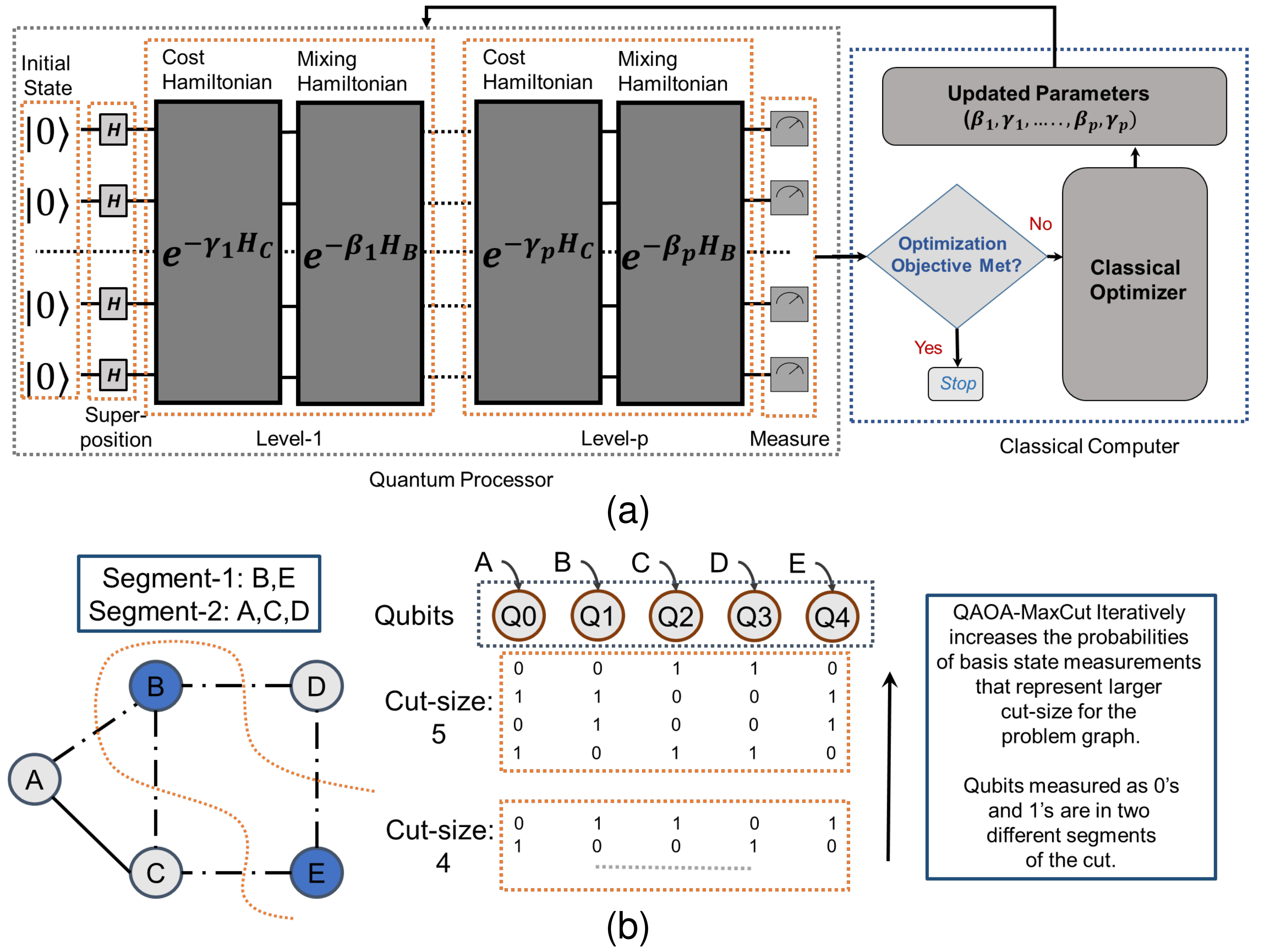}
 \end{center}
 \vspace{-1em}
 \caption{(a) Steps of QAOA for solving combinatorial optimization problems; (b) QAOA illustration for solving the maximum-cut (MaxCut) problem.} \label{fig:qaoa}
 \vspace{-4mm}
\end{figure}

\begin{figure*} [] 
\vspace{-1em}
 \begin{center}
    \includegraphics[width=0.95\textwidth]{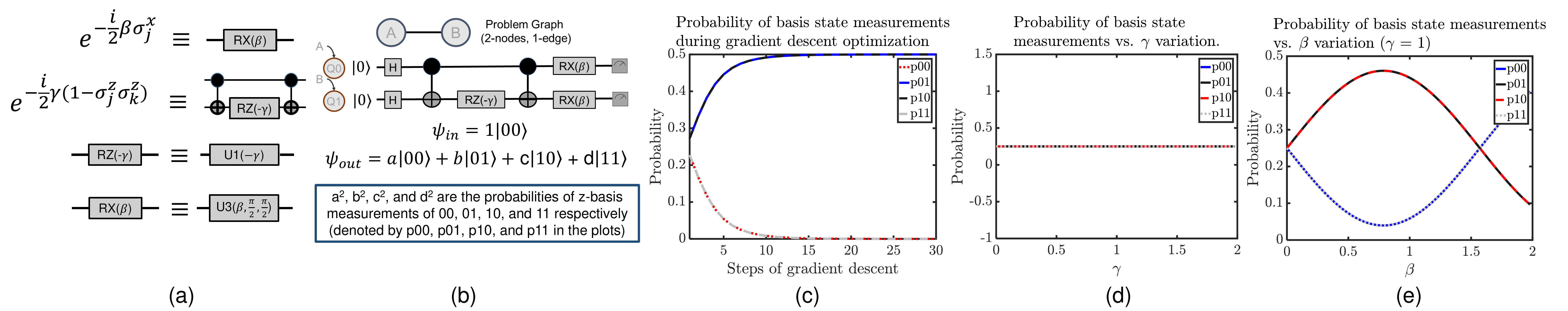}
 \end{center}
 \vspace{-1em}
 \caption{(a) Components of QAOA-MaxCut cost and mixing Hamiltonians and their circuit decompositions; (b) MaxCut problem formulation of a two-node, single-edge graph; (c) changes in the pure state measurement probabilities during the optimization procedure for the two-node graph QAOA-MaxCut problem; (d) pure states measurement probabilities after the cost Hamiltonian operation; (e) pure states measurement probabilities after the mixing Hamiltonian operation.} \label{fig:probability}
 \vspace{-4mm}
\end{figure*}

\subsection{QAOA for Combinatorial Optimization}
QAOA is a quantum-classical hybrid algorithm \cite{farhi2014quantum} which can tackle combinatorial optimization problems. In QAOA, each of the binary variables in the target $C(z)$ is represented by a qubit. The classical objective function ($C$($z$)) is converted into a quantum problem Hamiltonian by promoting each binary variable ($z_i$) into a quantum spin $\sigma_i^z$: $H_C$ = $C$($\sigma_1^z$, $\sigma_2^z$,...., $\sigma_N^z$). After initializing the qubits in the state $\ket{+}^{\otimes N}$, the problem (i.e., cost) Hamiltonian and a mixing Hamiltonian ($H_B$ = $\sum_{j=1}^{N}$ $\sigma_j^x$) are applied repeatedly (p times for a p-level QAOA) with a controlled duration to generate a variational wavefunction: $\ket{\psi_{p(\gamma,\beta)}}$ = $e^{-i\beta_pH_B}e^{-i\gamma_pH_C}$ .... $e^{-i\beta_1H_B}e^{-i\gamma_1H_C}\ket{+}^{\otimes N}$. Here, $\gamma_1$...$\gamma_p$ and $\beta_1$...$\beta_p$ variables denote the duration's of the applied problem Hamiltonian and the mixing Hamiltonian in p different levels of the QAOA circuit respectively. With optimal values of the control parameters, the output of the QAOA instance is sampled many times and the classical cost function is evaluated with each of these samples. 
The sample measurement that gives the highest cost is taken as the solution \cite{farhi2017quantum}. Currently, there exists multiple schools of thoughts for determining the optimal values of the control parameters.

\textbf{Single Instance Optimization:} The control parameters ($\gamma,\beta$) are optimized for a given instance of the problem (presented as an approach in the original QAOA \cite{farhi2014quantum}). This introduces a potentially expensive optimization step for every query. For smaller p-values (e.g. p = 2/3, etc.), the classical optimizer needs to optimize a small number (2p) of parameters (e.g. 4/6 etc.) for the problem instance. 
Starting with a random set of values of the control parameters, we then determine the expectation value of $H_C$ in the variational state $E_p(\gamma,\beta)$ = $\bra{\psi_{p(\gamma,\beta)}}H_C\ket{\psi_{p(\gamma,\beta)}}$. A classical optimizer iteratively updates these variables ($\gamma,\beta$) so as to maximize $E_p(\gamma,\beta)$. A figure of merit ($FOM$) for benchmarking the performance of QAOA is the approximation ratio $r$ = $\frac{E_p(\gamma,\beta)}{C_{max}}$ or its conjugate (1-$r$) \cite{crooks2018performance}.

In a practical QAOA application with a gate-based quantum computer, the chosen qubits (representing the binary variables of the classical cost function) are prepared in the superposition state $(\frac{\ket{0}+\ket{1}}{\sqrt{2}})$ by applying Hadamard gates on each of these qubits as shown in Fig. \ref{fig:qaoa}(a). The problem Hamiltonian and the mixing Hamiltonian are decomposed into parameterized quantum circuits (PQC) with the native gates of the target hardware where the parameters are used to control the duration of the applied Hamiltonians. For a p-level QAOA, the gates in the PQC with current parameter values are executed sequentially and the output of the quantum processor is measured in the basis state many times to get a distribution. 

Each QAOA circuit measurement in the basis state generates a candidate solution for the combinatorial optimization problem. The average value of the classical cost function over a finite number of measurements can also be an estimate of the expectation value of $H_C$ \cite{zhou2018quantum, brandao2018fixed}. A classical optimizer then updates the parameters of the PQC to maximize $E_p(\gamma,\beta)$. The measurements and associated cost values are saved during the optimization procedure. At the end of the procedure, the measurement associated to the largest cost can be taken as the approximate solution \cite{guerreschi2018qaoa}. 

\textbf{Batch Optimization:} In \cite{wecker2016training}, a batch-training approach was introduced where the ($\gamma,\beta$) values are determined through the training of a batch of problem instances. The authors showed that the trained parameters produces a state with high overlap with the optimal state for both the training instance set and a randomly generated test-set.

\textbf{Analytical Approach:} In the original QAOA algorithm \cite{farhi2014quantum}, the authors showed that if p does not grow with N (fixed p algorithm) and each bit is involved in no more than a fixed number of clauses, then there is an efficient classical calculation that determines the angles that maximize the expectation value of the cost Hamiltonian.

\textbf{Brute-force:} Parameter values can be determined through a brute-force search. The entire solution space is discretized and the parameter values that result in the highest expectation value of the cost (/problem) Hamiltonian are taken as the optimal set of control parameters. This approach is certainly unsuitable for QAOA instances with larger p-values.

Noisy qubits of a practical quantum computer can affect the performance of QAOA for all of these approaches. However, in this article, we only analyze the performance of the single instance QAOA optimization for the MaxCut problem which is described in the following Section.
\vspace{-2mm}

\subsection{QAOA at Higher Depths}

QAOA can be thought of as a quantum Hilbert space exploration procedure where the goal is to reach a quantum state with sufficient support to the optimal solution for a given combinatorial optimization problem. The control parameters ($\gamma_1$...$\gamma_p$ and $\beta_1$...$\beta_p$) are varied to explore the Hilbert space. Ideally, we are able to explore all the spaces that are reachable for the p-level QAOA instance with a (p+1) level QAOA instance. For example, a QAOA circuit with p=1 produces the quantum state with the highest possible support to the optimal solution for ($\gamma_{1opt}$,$\beta_{1opt}$). Ideally, p=2 QAOA instance will be able to produce the same state with ($\gamma_{1opt}$,$\beta_{1opt},\gamma_{2} = 0, \beta_{2} = 0$). Furthermore, it is possible for the p=2 QAOA instance to explore more spaces where quantum states with higher support to the optimal solution may reside. However, our analysis indicate that noise could alter the solution space from the ideal case, and the alteration increases with the higher p as more gates add more noise. 

\section{Solving MaxCut with QAOA}

\subsection{MaxCut Problem Statement}
MaxCut problem can be described as the following: given a graph $G$ = ($V$, $E$) with nodes $V$ and edges $E$, find a subset $S$ $\in$ $V$ such that the number of edges between $S$ and its complementary subset is maximized. 
Finding an exact solution of MaxCut is NP-hard \cite{karp1972reducibility}, however, there are efficient polynomial time classical algorithms that find an approximate answer within some fixed multiplicative factor of the optimum \cite{papadimitriou1991optimization}. In a classical setup, if the nodes of a target $N$-node graph are represented by the binary variables \{$z_1$,$z_2$,....,$z_N$\}, a MaxCut solving procedure maximizes following cost function: $\frac{1}{2}$ $\sum_{(i,j) \in E}^{}$ $C_{ij} (1 - z_{i}z_{j}$) where $C_{ij}$ = 1 if the nodes are connected and 0 otherwise (unweighted graph). A given graph can have many MaxCut solutions. A 5-node graph with a MaxCut solution is shown in Fig. \ref{fig:qaoa}(b).

\subsection{MaxCut-QAOA Formulation}
To solve the MaxCut problem with QAOA, we first convert the classical cost function into a cost Hamiltonian ($H_C$) by replacing the binary variables with Pauli-Z operations: $\frac{1}{2}$ $\sum_{(i,j) \in E}^{}$ $C_{ij} (1 - \sigma_{i}^{z}\sigma_{j}^{z}$). In this work, we have used following mixing Hamiltonian ($H_B$) \cite{farhi2014quantum}: $\frac{1}{2}$ $\sum_{i = 1}^{N}$ $\sigma_{i}^{x}$. For a p-level QAOA, a N-qubit quantum system is evolved with $H_C$ and $H_B$ p-times to find a MaxCut solution of a N-node graph with controlled durations: 
$e^{-i\beta_pH_B}e^{-i\gamma_pH_C}$ .... $e^{-i\beta_1H_B}e^{-i\gamma_1H_C}\ket{+}^{\otimes N}$. Each of the Pauli-X interactions in the mixing Hamiltonian can be implemented with a single one-qubit gate and each of the two-qubit ZZ interactions in the cost Hamiltonian can be implemented with two CNOT gates and a local single-qubit gate as shown in Fig. \ref{fig:probability}(a) \cite{crooks2018performance}. CNOT is a native gate of IBM quantum computers and the parametric RX($\beta$) and RZ($-\gamma$) operations can be realized using the available parametric U3 and U1 gates as shown in Fig. \ref{fig:probability}(a).

\begin{figure*} [!ht] 
\vspace{-1em}
 \begin{center}
    \includegraphics[width=0.9\textwidth]{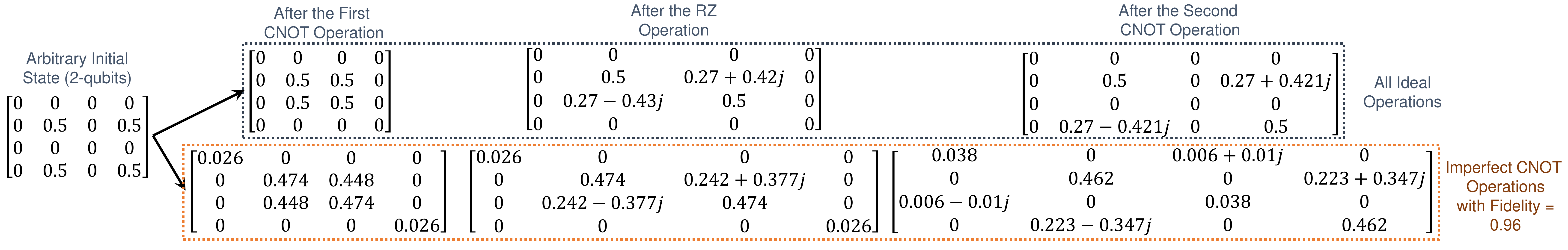}
 \end{center}
 \vspace{-1em}
 \caption{Infidelity in evolving the system with the cost Hamiltonian due to gate errors.} \label{fig:error_costH}
 \vspace{-4mm}
\end{figure*}

\subsection{Ideal Operation}
During the hybrid QAOA optimization procedure, the probabilities of the basis state measurements that represent larger cut-size for the given graph becomes larger with the increase in the expectation value of the cost Hamiltonian (Fig. \ref{fig:qaoa}(b)). To demonstrate the process, we have shown the probabilities of all four basis state measurements (for 2-qubits) during a MaxCut-QAOA optimization procedure (using gradient descent) for a 2-node graph with a single edge (Fig. \ref{fig:probability}(b)) in Fig. \ref{fig:probability}(c). Here, the $\ket{10}$ and $\ket{01}$ qubit assignments are the solutions to the given problem instance. 

When the system is evolved with the cost Hamiltonian, none of the basis state probabilities (amplitudes) are changed as shown in Fig. \ref{fig:probability}(d). However, it entangles the qubits by separating their phases. 
Hence, the step is also called \textit{phase separation} \cite{hadfield2017quantum}. The mixing Hamiltonian mixes amplitudes between the computational basis states. The entaglement ensures that a sub-space with higher probabilities of the feasible solutions in the quantum Hilbert space can be explored during the mixing step. For instance, after applying a constant phase factor ($\gamma$=1), it is possible to explore the sub-space with higher probabilities of the solutions ($\ket{01}$/$\ket{10}$) through amplitude mixing (by $\beta$ variation) as shown in Fig. \ref{fig:probability}(e).

\subsection{Non-ideal Operation}
\subsubsection{Imperfect Gate Operations}
Gate noises in practical quantum hardware changes the solution space for QAOA. The ZZ-interaction in the cost Hamiltonian which can be implemented with 2 CNOT and 1 RZ operations as shown in Fig. \ref{fig:probability}(a) changes the amplitudes of the basis states in practical hardware due to gate imperfections. The first CNOT gate either flips the amplitude of the target qubit or keeps it the same based on the amplitude of the control qubit. The RZ operation creates the desired phase separation while the last CNOT operation is supposed to restore the amplitude of the target qubit. However, due to erroneous gate operations, this restoration process irrecoverably changes the amplitudes of the qubits involved which ultimately changes the solution space for the subsequent mixing step. The phase difference between the qubits also deviates from the desired values during this process. 

To better illustrate the phenomena, we have shown the impact of these operations on the quantum state (denoted by a $4\times4$ density matrix) of a two-qubit system in Fig. \ref{fig:error_costH}. We have assumed the CNOT operations are erroneous in this example with Fidelity = 0.96 ($RZ() = U1()$ is a virtual gate in IBMQX without any error). In the ideal scenario, the diagonal elements of the output density matrix are exactly similar to the initial state density matrix values (indicating no change in the probabilities of different basis state measurements). However, when we consider imperfect gate operations, the output density matrix diagonal elements values are (0.038, 0.462, 0.038, 0.462) which are significantly different from their intended values (0, 0.5, 0, 0.5). The problem becomes severe for larger and realistic cost Hamiltonians with many imperfect gate operations e.g., for a graph with 100 edges, the fidelity of the cost Hamiltonian operation will be approximately $(1-e)^{200}$ where $e$ is the infidelity of a single CNOT operation. Even for a quantum computer with CNOT gate Fidelity = 0.99, the cost Hamiltonian operation will have a fidelity of approximately 0.366. In simple words, when we evolve the quantum system with the cost Hamiltonian, the change in the quantum state will be nowhere near to the intended. 

\begin{figure} [hb] 
\vspace{-1em}
 \begin{center}
    \includegraphics[width=0.43\textwidth]{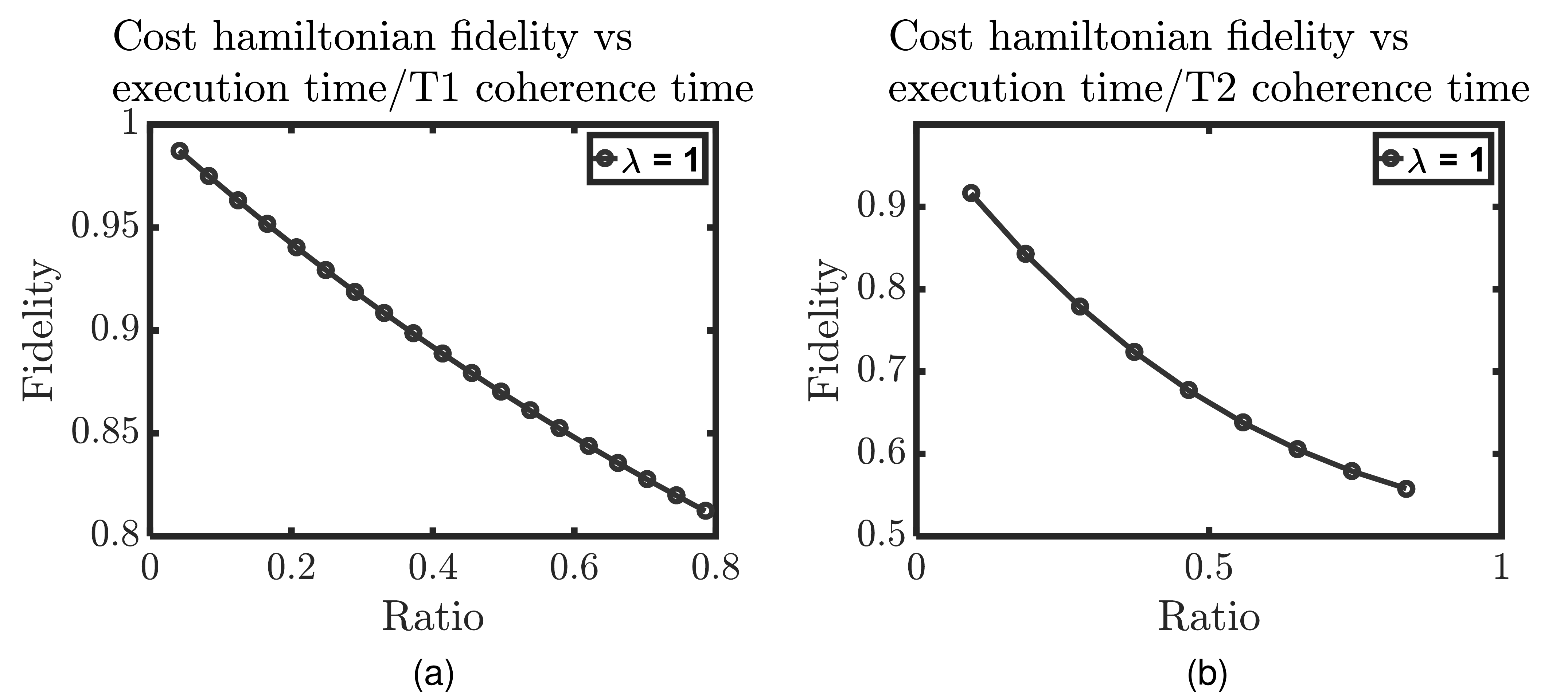}
 \end{center}
 \vspace{-1em}
 \caption{Impact of the circuit delay (or depth) on the fidelity of the cost Hamiltonian due to finite coherence time for the two-node, single-edge graph for (a) T1 and (b) T2 times of the modeled QC (discussed in Section \ref{PNQ}).} \label{fig:fid_decoherence}
 \vspace{-3mm}
\end{figure}

\subsubsection{Decoherence}
Similar arguments hold for decoherence induced errors. If the operation time for the cost Hamiltonian is not considerably shorter than the coherence times, the output state after the cost Hamiltonian operation will be extremely erroneous. To illustrate this, we have shown the infidelity of the cost Hamiltonian for the two-node graph against the ratio of the execution time and the coherence times (by varying the T1 and T2 coherence times and keeping the gate operation times fixed) in Fig. \ref{fig:fid_decoherence}(a) \& (b). We have assumed the qubits are in a superposition state before applying the cost Hamiltonian and the phase factor ($\lambda$) is set to 1. The level of error incurred due to relaxation/dephasing will also depend on the intermediate states of the quantum system. However, the trend shown in Fig. \ref{fig:fid_decoherence} will be similar for any arbitrary initial states and phase.

\subsubsection{QAOA Solution Space}
\begin{figure} [hb] 
\vspace{-1em}
 \begin{center}
    \includegraphics[width=0.42\textwidth]{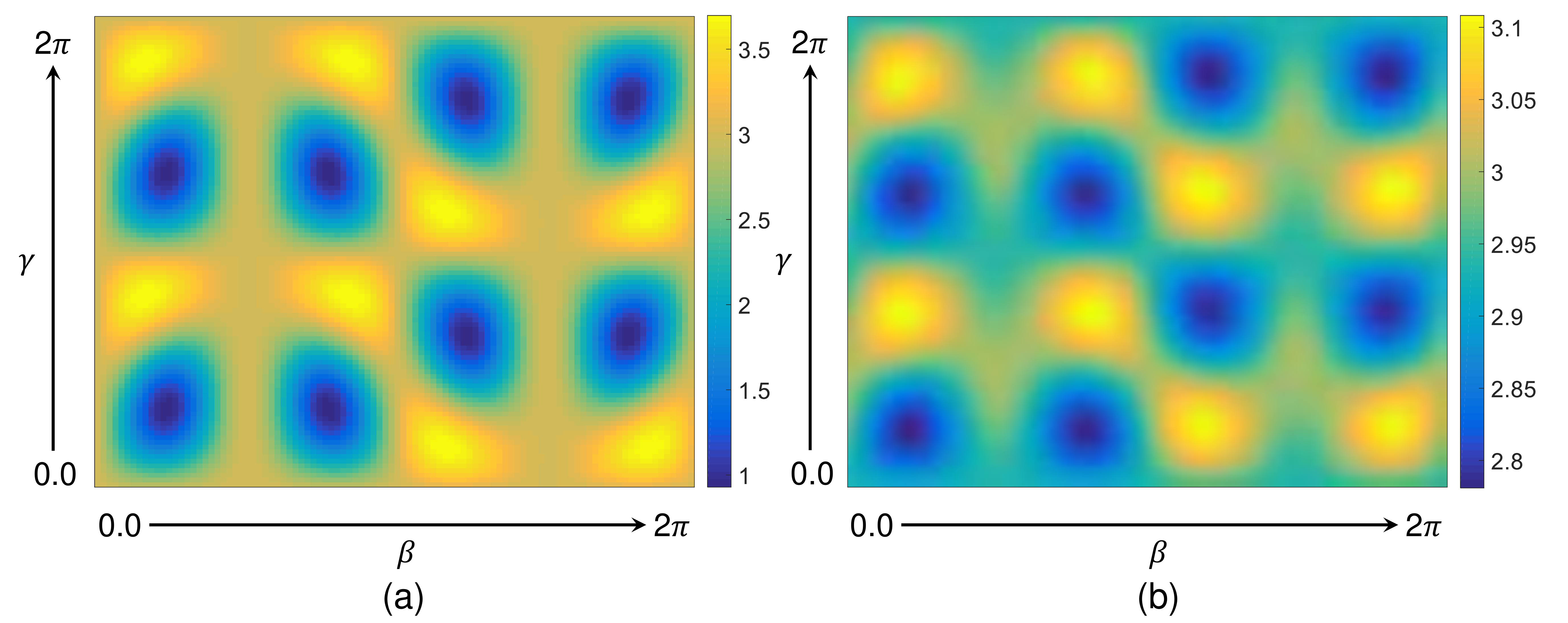}
 \end{center}
 \vspace{-1em}
 \caption{QAOA solution space for the 4-node 3-regular yutsis graph (a) noiseless, and (b) with noises.} \label{fig:space}
 \vspace{-4mm}
\end{figure}

To explore the cumulative impact of the error sources, we have simulated the entire solution space ($\gamma$/$\beta$ = 0 to 2$\pi$) for the 4-node yutsis problem instance (p=1) with noiseless and noisy qubits based on the noise characteristics of the modeled QC (discussed in Section \ref{PNQ}) and the expectation values of the cost Hamiltonian are shown in Fig. \ref{fig:space}(a) and (b). Note that, for the noiseless condition, it is possible to get a set of control parameters to maximize the expectation value to 3.7. Therefore, if we sample the output of the variational circuit (for this optimal control parameters), it is highly likely to sample the basis states that represent the best solution (MaxCut = 4) for the problem. However, for the noisy qubits, the maximum expectation value obtained is 3.1. Therefore, it is less likely to sample the basis state from the circuit output that represent the best solution with the optimal control parameters. 

In the next Section, we analyze the impact of various noise sources on the QAOA performance for larger MaxCut instances. 

\begin{figure*} [!ht] 
\vspace{-1em}
 \begin{center}
    \includegraphics[width=0.98\textwidth]{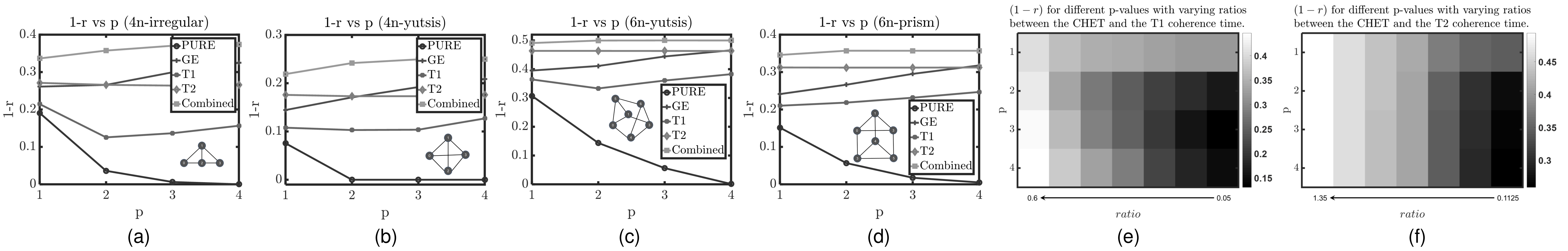}
 \end{center}
 \vspace{-1em}
 \caption{FOM $(1-r)$ values of the global optimization procedure with and without noise sources for (a) 4-node random irregular graph, (b) 4-node yutsis, (c) 6-node yutsis, (d) 6-node prism unweighted 3-regular (u3R) graphs; Optimal p-value dependency on the ratio of the cost Hamiltonian execution time and the (e) T1 coherence time, and (f) T2 coherence time for the 6-node yutsis graph.} \label{fig:error}
 \vspace{-1em}
\end{figure*}

\section{QAOA Performance with Noisy Qubits} \label{PNQ}
\subsection{Simulation Setup}
\subsubsection{Target Problems}
Simulation of quantum systems in classical computers (beyond a few qubits) is resource-intensive and require prohibitively large amount of memory and computational power \cite{guerreschi2018qaoa}. Therefore, we have confined our simulations to graphs with 6-nodes or fewer and selected 3 unweighted 3-regular graphs (u3R) for our analysis with 4 nodes (4n-yutsis), and 6 nodes (6n-yutsis and 6n-prism) and an irregular 4 node graph (4n-irregular). 

\subsubsection{Modeled Quantum Computer} \label{MQC}
To eliminate the impact of the coupling constraints on the QAOA performance, we have modeled a 6-qubit fully connected quantum computer (QC) where two-qubit operation (CNOT) is allowed between any two qubits. The modeled QC supports single-qubit U1, U2, and U3 operations along with CNOT similar to IBMQX4. All the qubits in the model are considered identical in terms of their quality metrics. The T1-relaxation and T2-dephasing time for each of the qubits are taken as average values of these parameters in IBMQX4 based on the reported calibration data collected over a 50-day period ($45\mu s$ and $20\mu s$ respectively) \cite{IBMQ}. This is done to simplify the analysis. The single-qubit and two-qubit gate-errors are also modeled in a similar fashion ($1.5 \times 10^{-3}$ and $4\times10^{-2}$). The U1 (virtual gate), U2, U3, and CNOT gate operation times are taken as 0, 60ns, 120ns, and 720ns respectively \cite{IBMQ}.

\subsubsection{Classical Optimizer}
A variety of classical optimizers have been used for QAOA circuits, including Nelder-Mead, Monte-Carlo, Quasi-Newton, Gradient Descent, Bayesian Methods, etc. The distribution of the initial parameters can have an impact on the training performance (for local optimizers) \cite{crooks2018performance}. If the distribution of these initial parameters is overly broad, then the output quantum state of the circuit (Ansatz) is essentially random and hard to optimize. Moreover, local-optimizer such as Nelder-Mead may get stuck in a local optima during the optimization which can also affect the QAOA performance. In this article, we analyze the behavior of QAOA performance in presence of noise from a circuit perspective. Hence, to minimize the impact of the classical optimizer on the QAOA performance (e.g. being stuck in a local optima), we have used a global optimizer named differential-evolution from SciPy-optimize library \cite{storn1997differential}. We restrict our optimization domain to $\beta_i$ $\in$ [0,$\pi$], $\gamma_i$ $\in$ [0,2$\pi$] as in \cite{farhi2014quantum}.

\subsubsection{Circuit Execution Times}

In our modeled QC, the gates in the given quantum workload are executed one at a time. The total execution time for any given quantum circuit (also called circuit latency) is the summation of the gate-operation time of the individual gates in the workload. For any p-level QAOA circuit instance, the circuit representation of the cost and the mixing Hamiltonians are executed p-times. Therefore, the entire circuit execution time is proportional to the depth of cost and mixing Hamiltonian execution times. For instance, the QAOA circuit instance for the 4n-yutsis problem graph has 12 CNOT gates (6 edges), 6 U1 gates for the RZ(-$\lambda$) gates, 4 U3 gates for the RX($\beta$) operations, and 4 U2 gates for the Hadamard operations at initial state preparation representing a total circuit execution time of $(12\times720+6\times0+4\times120+4\times60)ns$ or $9.36\mu s$ for p=1. The circuit execution time is $(9120\times2 + 4\times60)ns$ or $18.48\mu s$ for p=2. Note that, the cost Hamiltonian circuit execution time constitutes the major portion of the total QAOA circuit execution time. For instance, the cost Hamiltonian execution time for the 4n-yutsis graph is $(12\times720+6\times0)ns$ or $8.64\mu s$ ($\approx$90\% of the total circuit execution time for p=1).

\vspace{-2mm}
\subsection{Impact of Noises on QAOA}
\subsubsection{Relaxation}

We first analyze the QAOA performance with the reported relaxation (T1 coherence) time of our modeled QC. We optimize the control parameters for QAOA instances for the chosen graphs with p-values varying from 1 to 4. The figure of merits (FOM) of the optimization procedure, (1-r), are plotted in Fig. \ref{fig:error}. A small FOM value indicates a large approximation ratio for the MaxCut problem (i.e., a better solution). Note that, the performance benefit observed in noiseless condition ($PURE$ in Fig. \ref{fig:error}) with increasing p-values is not attainable when we consider relaxation errors, and this is true for all the graphs we have considered ($T1$ in Fig. \ref{fig:error}). For instance, we have received the FOM values of 0.308, 0.144, 0.056, and 0.001 for p = 1, 2, 3, and 4 respectively in the noiseless scenario for the 6n-yutsis graph after the optimization. In presence of relaxation errors, the FOM values are 0.364, 0.333, 0.361, 0.382 respectively which are considerably larger. Note that, the circuit operation time for the 6n-yutsis instances are $15.48\mu s$, $30.24\mu s$, $45.00\mu s$, and $59.76\mu s$ for for p = 1, 2, 3, and 4 respectively. The T1 time is $45\mu s$. In presence of relaxation error, the FOM values decrease from p=1 to p=2. For these cases, the circuit operation times are smaller than the T1 time. However, for larger p-values (p=3 and 4), we have observed noticeable increase in the FOM values. For p=3, the circuit execution time is similar to T1 time and for p=4, it exceeds the T1 time. 

\begin{figure*} [!ht] 
\vspace{-1em}
 \begin{center}
    \includegraphics[width=0.98\textwidth]{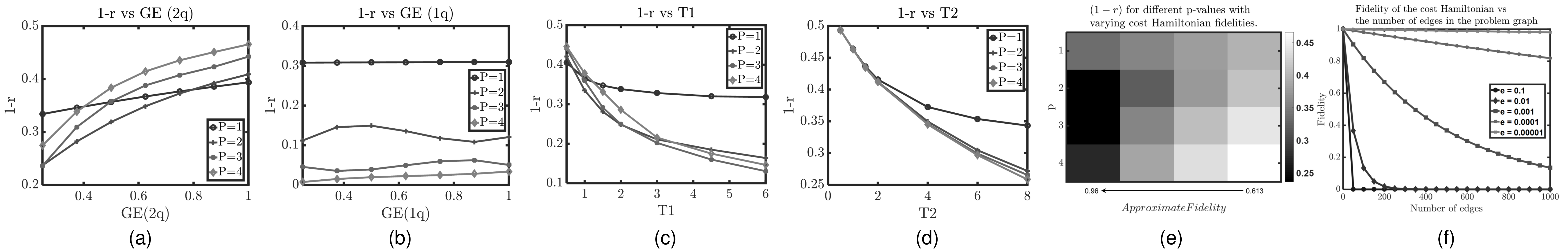}
 \end{center}
 \vspace{-1em}
 \caption{FOM $(1-r)$ values of the global optimization procedure for the 6-node yutsis graph with (a) two-qubit gate (CNOT) error, (b) single-qubit gate (U2/U3) error, (c) T1, and (d) T2 variations; (e) Optimal p-value dependency on the approximate fidelity of the cost Hamiltonian for the 6-node yutsis graph; (f) approximate fidelity of the cost Hamiltonians with varied number of edges of the problem graphs and two-qubit gate (CNOT) error rates.} \label{fig:sweep}
\end{figure*}

\textbf{Optimal p-bound for finite T1 time:} Generally, if the qubit relaxation time is finite, QAOA with high p-values will give better FOM values if the circuit execution time is considerably smaller than the T1 time. To validate this, we swept T1 for the 6n-yutsis QAOA instances (0.5xT1 to 6xT1 where T1 is the reported relaxation time of our modeled QC - $45\mu s$) and the corresponding FOM values are shown in Fig. \ref{fig:sweep}(c). For 1xT1, p=2 provides the best FOM value. With larger T1 (e.g., 3xT1), p=3 provides the best FOM. The results indicate that, for any target quantum hardware and given problem instance, there will be a finite bound to the optimum number of stages (p-value) which will be dictated by the ratio of the cost Hamiltonian execution time ($CHET$) and the T1 coherence time opposing to the general belief that larger QAOA instances (high p-values) will always give better solutions than the smaller ones. For the 6n-yutsis graph MaxCut problem, the FOM $(1-r)$ values for different p-values and the CHET/T1 coherence time ratios are shown in Fig. \ref{fig:error}(e).

\begin{figure*} [!ht]
\vspace{-1em}
 \begin{center}
    \includegraphics[width=0.9\textwidth]{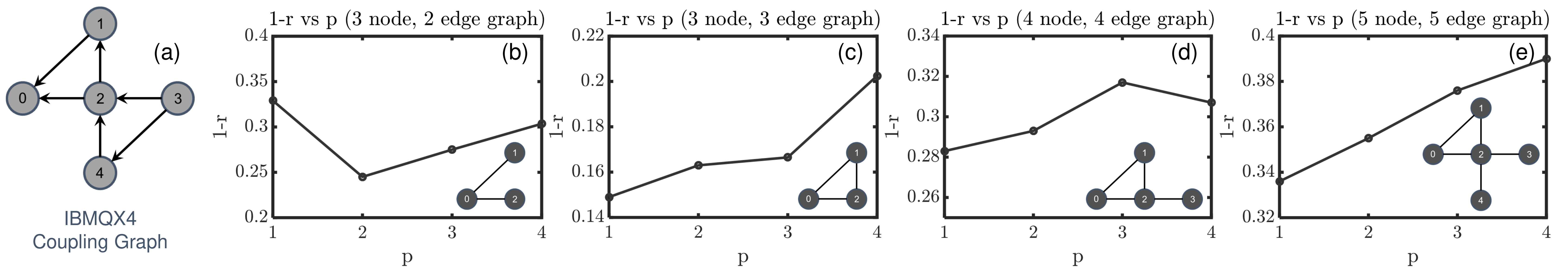}
 \end{center}
 \vspace{-1em}
 \caption{FOM $(1-r)$ values for different problem graphs and p-values on IBMQX4. The coupling graph of IBMQX4 is shown in (a). Optimal p-value was found to be 1 for the graph instances in (c), (d), and (e). For the smallest graph instance, optimal-p value was found to be 2 in (b).} \label{fig:hw}
 \vspace{-3mm}
\end{figure*}

\subsubsection{Dephasing}
We have performed the global optimization for the chosen QAOA instances in presence of dephasing errors (T2 coherence time of our modeled QC) and the results are shown in Fig. \ref{fig:error}. We note practically no change in the FOMs for any of the chosen graphs with varying p-values ($T2$ in Fig. \ref{fig:error}). This is because the circuit execution times for all the instances are either extremely close to or exceeding the T2 coherence time of the modeled QC ($20\mu s$). However, in all the cases, the FOM's are substantially degraded than the noiseless cases.  

\textbf{Optimal p-bound for finite T2 coherence time:} Similar to the relaxation error, dephasing puts a finite bound to the optimal p-value for any problem instance and a target quantum hardware. To validate this, we swept the T2 coherence times from 0.5xT2 to 8xT2 (where T2 is the reported T2 dephasing time of the modeled QC - $20\mu s$) for the 6n-yutsis problem graph and the corresponding FOM values are shown in Fig. \ref{fig:sweep}(d). For dephasing times up to 2xT2, there is no significant change between the FOMs for different p values. For 4xT2, p=2 performs significantly better than p=1. However, increasing p further for 4xT2 does not improve the solution noticeably. For larger dephasing time, 6xT2, the gap between the FOM's of p=2 and p=3 widens. The results indicate that, in presence of dephasing errors, increasing p-value will improve FOMs up to a certain point. Increasing p beyond that limit is unlikely to provide better solutions. Similar to the relaxation errors, the optimal p-value depends on the ratio of the cost Hamiltonian execution time ($CHET$) and the T2 coherence time. For the 6n-yutsis graph MaxCut problem, the FOM $(1-r)$ values for different p-values and the CHET/T2 coherence time ratios are shown in Fig. \ref{fig:error}(f).

\subsubsection{Gate Error}

The results for global optimization of the MaxCut instances in presence of the reported gate errors of our modeled QC are shown in Fig. \ref{fig:error}. We noted no performance improvement with increasing p-values in presence of gate errors for all the problem graphs. Rather, the FOM $(1-r)$ values are substantially degraded for higher-depth QAOA instances compared to the lower ones ($GE$ in Fig. \ref{fig:error}). 

\textbf{Optimal p-bound for gate error:} The QAOA performance with larger p-values is most likely dependent on the fidelity of the cost Hamiltonian operation. The approximate fidelity of the cost Hamiltonian of the chosen graph instances in Fig. \ref{fig:error} are 0.72, 0.61, 0.48, and 0.48 respectively (for the reported gate errors of our modeled QC). To validate our assumption, we have swept the two-qubit and single-qubit gate error rates separately (from 0.25xGE to 1xGE where the GE is the reported gate-error rate of our modeled QC) for the 6n-yutsis problem graph and the results are shown in Fig. \ref{fig:sweep}(a) \& (b). Note that, the single-qubit gate errors do not show much impact on the QAOA performance with varying depths. The FOM $(1-r)$ rather follows the expected trend - lower $(1-r)$ for larger p as shown in Fig. \ref{fig:sweep}(b) for all the noise levels (0.25xGE to 1xGE). The reason behind this is that the reported single-qubit gate errors are extremely small to have any noticeable impact on the cost Hamiltonian fidelity. However, when we swept the two-qubit gate error rates, we have found that at smaller error levels, larger-depth QAOA instances have provided better FOM values as evident from Fig. \ref{fig:sweep}(a). For instance, at 1xGE error level, p=1 provides the best FOM. At 0.6xGE error level, p=2 provides the best FOM. The FOM values for the 6n-yutsis graph instance against the approximate fidelity of the cost Hamiltonian is shown in Fig. \ref{fig:sweep}(e) which also corroborates our claim on the dependence of the optimal p-depth on the fidelity of the cost Hamiltonian for any given QAOA instance. 

Note that, the optimal p-bound due to gate errors is quite significant. It indicates that, even if we are able to realize qubits with infinite coherence time, the performance improvement with larger p-values will be unlikely due to the gate errors for larger and realistic problems. In Fig. \ref{fig:sweep}(f), we have shown the approximate fidelity ($(1-e)^{2n}$) of the cost Hamiltonian with varying number of edges ($n$) in the problem graph and different noise levels ($e$) for the CNOT operation. For the 6n-yutsis problem graph, we have found no performance improvement beyond p=1 when the approximate fidelity of the cost Hamiltonian was 0.613 (Fig. \ref{fig:sweep}(e)). With 40x improvement in the CNOT gate error (from the reported error value of the modeled QC (0.04) to 0.001), the approximate fidelity of a realistic problem graph with a 1000 nodes will be around 0.135 (Fig. \ref{fig:sweep}(f)). A performance improvement for such problem instances and noise levels with p-values larger than 1 is implausible.

\subsubsection{Compound Error}

The interplay of all the error sources (relaxation, dephasing, gate-errors) creates a solution space for QAOA which can be significantly different from the ideal one (Fig. \ref{fig:space}). Hence, the behavior of the QAOA algorithm may differ from its expected when it is practically implemented in a hardware. The results of QAOA instances with all the error sources combined are shown in Fig. \ref{fig:error} ($Combined$). Note that, contrary to the ideal cases where a performance improvement is expected for QAOA instances with higher p-values, we observe a significantly degraded performance. For the current noise values, any performance improvement beyond p=1 is unlikely. However, with better qubits in the horizon, the optimal p-values can shift.

\subsection{Hardware Validation}\label{sec:hw}

To validate our observations on the optimal p-bound in noisy qubits, we have executed the QAOA circuit instances for 4 different graphs with their optimal control parameter values (found through the global optimization procedure with a modeled 5-qubit QC identical to IBMQX4 in Fig. \ref{fig:hw}(a), noise is modeled based on the reported calibration data) 
on IBMQX4 \cite{IBMQ}. The graphs are chosen carefully so that the QAOA circuit instances are already nearest-neighbour coupling compliant (to eliminate swap-based qubit allocation that can affect the QAOA performance). For any given circuit and p-value, the circuit output has been sampled 8192 times and the associated classical cost function mean value for these samples have been taken as the expectation value of the cost Hamiltonian. The FOM $(1-r)$ values for all the graph instances and p-values are plotted in Fig. \ref{fig:hw}. Note that, for the circuit with the lowest number of gates among the QAOA instances (Fig. \ref{fig:hw}(b) - depth or the number of gates in the QAOA instances are proportional to the number of edges for the chosen graphs), $(1-r)$ decreased from p=1 to p=2. After p=2, the $(1-r)$ values increased with the increase in the p-value. For this problem instance, p=2 is the optimal QAOA-depth. For all other graph instances, p=1 was found to be the bound for optimal p. These results agree with our simulation based analysis.

\section{Conclusion}

We analyzed the performance of QAOA-MaxCut with realistic noise attributes of superconducting qubits. Our results indicate that the optimal p-value for any QAOA single instance optimization procedure will be bounded by the qubit quality metrics (/noise characteristics) of any target hardware. 
For the test cases considered in this work, the highest value of optimal p is 2 (experimental). For most of the practical size problems with existing error-rates, any higher value beyond p = 1 does not guarantee improved performance. 
Although, the bound on the optimal p-value due to finite coherence time is instinctive, a similar bound due to the gate error is significant. The results indicate that, even if we are able to realize qubits with infinite coherence time, the optimal depth (to begin with) for any QAOA single instance optimization procedure can be bounded by the gate error rates of the target hardware.

\bibliographystyle{IEEEtran}

\bibliography{IEEEabrv,biblio}

\end{document}